\documentclass[12pt]{iopart}

\usepackage{graphicx}
\usepackage{color}

\begin{document}

\title{Fibre-integrated noise gating of high-purity heralded single photons}

\author{Robert J. A. Francis-Jones and Peter J. Mosley}

\address{Centre for Photonics and Photonic Materials, Department of Physics, University of Bath, Bath, BA2 7AY, United Kingdom}

\ead{r.j.a.francis-jones@bath.ac.uk}

\begin{abstract}
We present an all-fibre source of high-purity heralded single photons with an integrated conditional optical gate that reduces uncorrelated noise by almost an order of magnitude. Generating photon pairs by four-wave mixing in photonic crystal fibre, we observe with the noise gate active a factor of 7 reduction in the rate of single counts in the heralded channel with no measurable drop in coincidence count rate. In contrast to electronic post-selection of coincidence events, the real reduction in the flux of unwanted photons is beneficial for example to avoid bleaching light-sensitive samples or in generating entangled states.
\end{abstract}

\maketitle

\section{Introduction}
Heralded single photon sources are vital for delivering the nonclassical states of light required in photonic quantum technologies~\cite{Kok2007Linear-Optical-Quantum-Computing}. Typically these sources involve photon-pair generation by intense laser pulses propagating through a nonlinear optical medium to create daughter photons known as the signal and idler. Often parametric downconversion (PDC) in a second-order nonlinear crystal is used~\cite{Burnham1970Observation-of-Simultaneity-in-Parametric}, however optical fibre sources based on spontaneous four-wave mixing (FWM) mediated by third-order nonlinearity can offer greater flexibility~\cite{Rarity2005Photonic-Crystal-Fiber-Source,Clark2011Intrinsically-Narrowband-Photon-Pair,Soller2011High-Performance-Single-Photon}. In particular, control of the waveguide dispersion in photonic crystal fibre (PCF) by changing the size and separation of the air holes in the cladding eases the production of photon-pairs directly in high-purity states -- a critical requirement for quantum-information applications. Although the third-order nonlinearity available is smaller than the second-order nonlinearity in crystals without inversion symmetry, much longer interaction lengths are possible enabling generation rates to remain high. Furthermore, with careful design of the fibre, both daughter photons can be generated in a single spatial mode~\cite{Birks1997Endlessly-Single-Mode}.

The drawback of generating photon pairs in a material without second-order nonlinearity is that unwanted third-order nonlinear effects can occur with a similar strength to that of FWM. Most importantly, spontaneous Raman scattering can populate the target frequencies with uncorrelated photons, particularly on the long-wavelength side of the pump~\cite{Lin2007Photon-Pair-Generation}. The impact of Raman noise can be mitigated to some extent by using dispersion-engineering to place the idler photons beyond the majority of the Raman gain~\cite{Smith2009Photon_Pair_Generation,Soller2010Bridging-Visible-and-Telecom} or by cooling the fibre cryogenically~\cite{Dyer2008High_Efficiency_Ultra}. In addition, correlation between the signal and idler modes can be used to gate desired events, either electronically through post-selecting coincident detection signals \cite{Pooser2012FPGA-Based-Gating}, or optically by feed-forward from the heralding detection to a fast optical switch~\cite{Brida2011Experimental-realization-of-a-low-noise}. The latter is more desirable as it reduces, rather than masks, the flux of unwanted photons delivered at the output of the source -- important for measurements that are limited by total photon budget~\cite{Matthews2016Towards_Pratical_Quantum, Sabines-Chesterking2016Sub-shot-noise-transmission-measurement}.

In this Letter we present a noise gate operating on high-purity heralded photons in a fully fibre-integrated architecture. Our source is dispersion-engineered to produce photon pairs without frequency correlation and as a result the photons cannot be subjected to narrowband spectral filtering. As they are also generated and routed entirely in guided modes it is especially difficult to achieve sufficient isolation of the photon pairs from the pump and discard Raman noise. We demonstrate the importance of fibre-integrated gating in limiting noise from uncorrelated photons and show that our gate reduces the level of uncorrelated single counts by a mean factor of 7.

\section{Source engineering for photon pair generation in PCF}

In order to deliver heralded photons in pure states, the key requirement is that photon pairs are created in a single pair of modes. However, energy and momentum conservation typically force pair generation into many correlated spectral modes; heralded photons are subsequently projected into an incoherent mixture by detecting their twin~\cite{Grice1997Spectral-Information-and-Distinguishability}. To avoid this, we designed the dispersion of our fibre so that the group velocities $v_j$ of the fields satisfy $v_{s} \leq v_{p} \leq v_{i}$, where $j = p,~s, ~\textnormal{and } i$ denote pump, signal and idler respectively, in order to generate signal and idler with minimised frequency correlation~\cite{Grice2001Eliminating-Frequency-and-Space-Time, Mosley2008Heralded-Generation-of-Ultrafast, Garay-Palmett2007Photon-Pair-State-Preparation}.

\begin{figure}
	\centering
    \includegraphics[width = 0.9\textwidth]{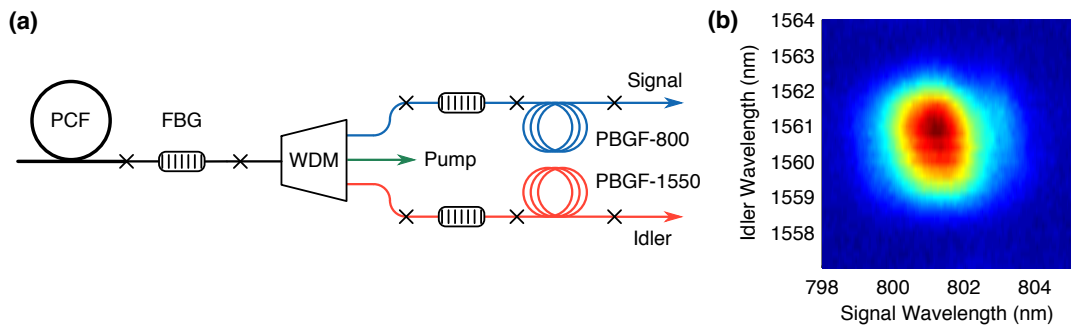}
    \caption{a.) Schematic of the photon-pair source. Photon-pairs were generated by FWM in a length of photonic crystal fibre. Fibre Bragg gratings (FBG) centred at the pump wavelength were used to reject the pump light and a wavelength division multiplexer (WDM) split the pump, signal and idler into three channels. Broadband filtering to isolate single-photon channels from the pump was applied using lengths of photonic bandgap fibre. (b) Joint spectral intensity distribution of the generated photon-pairs measured using stimulated emission tomography \cite{Francis-Jones2016All_Fibre_Multiplexed}.}
    \label{fig:Source_Schematic}
\end{figure}

Figure~\ref{fig:Source_Schematic} shows a schematic of the source. Photon pairs were produced at 800~nm and 1550~nm in a dispersion-engineered PCF, where the group velocity of the signal photon and pump were matched. The PCF was pumped at a repetition rate of 10\,MHz by an amplified modelocked 1064\,nm \emph{Fianium} FemtoPower 1060-PP fibre laser. The pump pulse duration was increased from 200\,fs to approximately 1\,ps in a 4f-grating spectrometer to optimise the pumping conditions for single mode photon-pair generation. A fibre Bragg grating (FBG) spliced onto the PCF was used to attenuate the pump light and a fibre wavelength division multiplexer split the light into three separate fibres: the residual pump, signal, and idler arms. In the signal and idler arms additional FBGs were included to increase the attenuation of the pump. Broadband filtering, to isolate the signal and idler photons from the majority of the noise whilst not affecting the joint spectral distribution, was achieved using high-loss wavelength regions in a photonic bandgap fibre (PBGF-800 and PBGF-1550). All the components were fusion spliced to form a completely fibre-integrated source that required no further alignment beyond coupling the pump light into the PCF.

Further details of our source are given in \cite{Francis-Jones2016All_Fibre_Multiplexed}, where we present measurements of the joint spectral intensity distribution and marginal second-order coherence. To determine the level of spectral correlation present between the signal and idler in the two-photon state we measured the joint spectral intensity, as shown in Fig.~\ref{fig:Switching}b, using stimulated emission tomography~\cite{Liscidini2013Stimulated_Emission_Tomography}. This placed an upper limit on the heralded single photon purity of 0.86; the marginal second-order coherence yielded a reduced-state purity of 0.71. We believe this to be limited by the quality of the pump spectrum and inhomogeneity in the PCF.

\section{Optical noise gating with feed-forward}

\begin{figure}
	\centering
    \includegraphics[width = 0.9\textwidth]{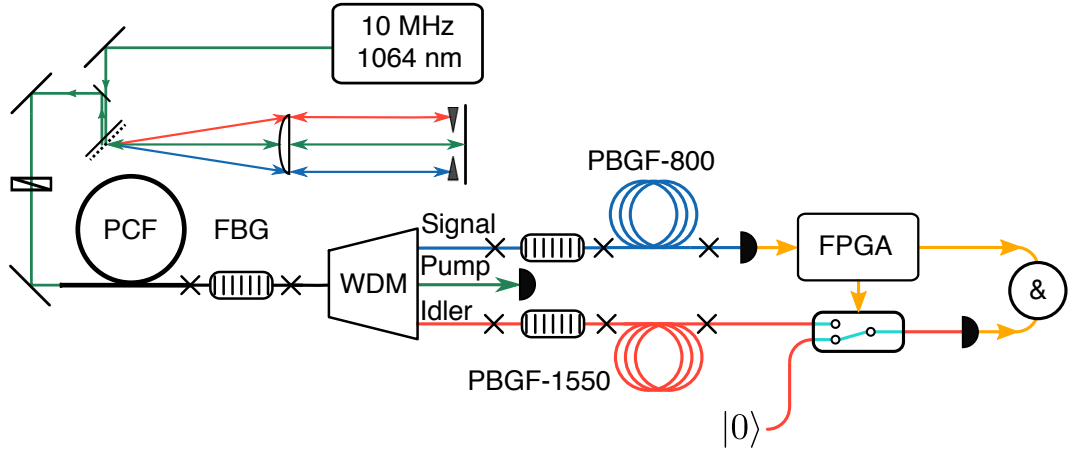}
    \caption{Schematic of the all-fibre source with noise gate. The heralding detector placed in the 800~nm signal arm fed forward to an FPGA to set the state of the fibre-coupled optical switch in the 1550~nm idler arm. The heralded idler photons are detected by an InGaAs APD and coincidences with the heralding detector are detected using custom FPGA-based coincidence logic.}
    \label{fig:Switching}
\end{figure}

The source was interfaced with a fast, low-loss $2 \times 1$ optical switch in the fully-spliced fibre architecture shown in Fig.~\ref{fig:Switching}a. Single mode (SM800) fibre was spliced onto the output of the signal channel for coupling to the silicon avalanche photodiode (APD) used as a heralding detector. In the idler channel, the output of the PBGF was spliced onto approximately 40~m of single mode fibre (SMF28) to delay the idler photon by around 200~ns, before being spliced to one input of the $2\times 1$ switch (insertion loss approximately 1 dB, rise time less than 100 ns).

The signal from the silicon APD was fed forward via a field programmable gate array (FPGA) to control the switch. In its default state, the switch connected the vacuum state input to the output; whenever a heralding detection occurred the FPGA set the switch to route the output of the delayed idler arm to the common output. In total, there was around 200~ns of latency in the heralding and switch electronics, to which the optical delay of the idler mode was matched to enable the switch to open prior to the arrival of the corresponding photon. The switch window was set to $\Delta t_{sw} = 100$~ns. Finally, the SMF-28 fibre switch output was connected to an InGaAs APD (ID Quantique ID210) and coincidence logic between the two detectors was carried out with another FPGA.

\section{Results}

We characterised the performance of the noise gate by measuring single and coincidence count rates as a function of pump power, with and without the noise gate activated. Count rates were recorded for periods of 61\,s and the data shown are the mean rates per second.

\begin{figure}
	\centering
    \includegraphics[width = 0.5\textwidth]{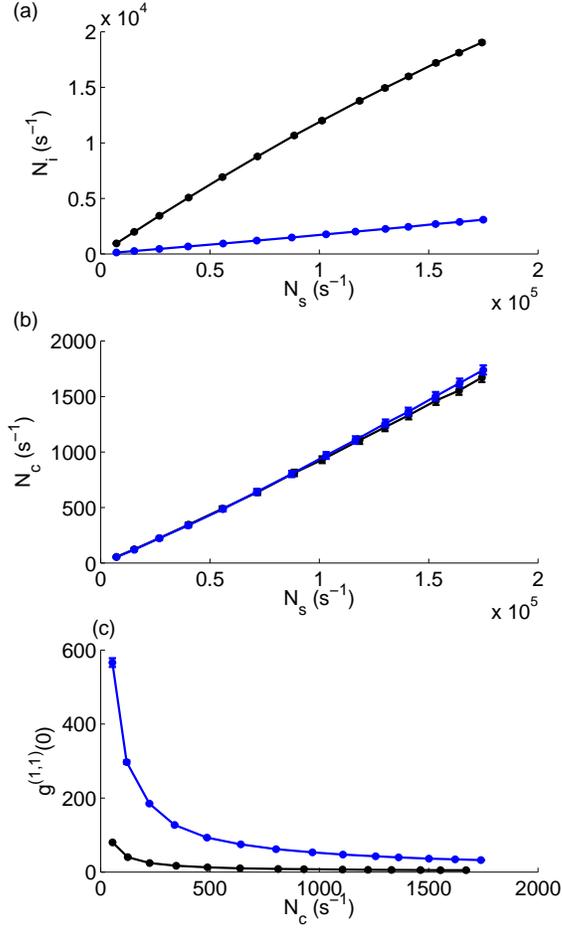}
    \caption{(a) Mean idler count rate and (b) coincidence count rate as a function of signal count rate. (c) Signal-idler cross correlation function, all plotted without (black) and with (blue) noise gating active. Error bars reflect Poissonian counting statistics and are often too small to be seen.}
    \label{fig:Combined_Results}
\end{figure}

The raw count rates are displayed in Fig.~\ref{fig:Combined_Results}a. We see that activating the noise gate drastically reduces the number of single counts at the idler detector $N_{i}$ while leaving the coincidence count rate unaffected. For example for an average pump power of approximately 0.2\,mW, when the noise gate is running idler single counts fall from 2000/s to 290/s with no measurable change in coincidence count rate. The mean reduction in idler singles as a result of the noise gate across the range of count rates investigated is 7.0(4); to within error there was no mean change in the coincidence count rates. Furthermore, we note that without the noise gate running the dependence of the idler count rate on the signal count rate is not linear above a herald count rate of approximately $10^5$/s; this is due to the onset of detector saturation as a result of the 10\,$\mu$s dead time of the InGaAs APD. The same behaviour is not observable with the noise gate on over the range of heralding count rates investigated here, suggesting that the noise gate significantly increases the range of count rates over which the system can be operated.

The improvement in performance wrought by the noise gate can be further elucidated by considering the strength of correlation between the output modes. We calculate the first-order cross-correlation between the signal and idler channels, $g^{(1,1)}$:
\begin{equation}
g^{(1,1)} = \frac{p_{s,i}}{p_s p_i}
\end{equation}
where $p_{s,i}$ is the probability of a coincidence event per time bin, and $p_s$ and $p_i$ are the corresponding probabilities of single events in the signal and idler channels. This is related to the measured count rates by: 
\begin{equation}
	g^{(1,1)} = \frac{N_{s,i} R_{p}}{N_{s} N_{i}}.
    \label{eq:xcorr}
\end{equation}
where $N_{s}$ and $N_i$ are the count rates per second of signal and idler detectors respectively, $N_{s,i}$ is the coincidence count rate, and $R_{p}$ is the pump repetition rate (10\,MHz). When considering photon-pair sources, $g^{(1,1)}$ is often referred to as the coincidence-to-accidental ratio (CAR); $g^{(1,1)} > 2$ implies that the signal and idler modes are more correlated than would be the case were they derived from a single-mode thermal state.


Figure~\ref{fig:Combined_Results}b shows $g^{(1,1)}$ as a function of the coincidence count rate, with and without the noise gate active. We see that at a fixed coincidence count rate the noise gate increases the cross-correlation over the entire range measured, also by a factor of approximately 7. With the noise gate active, the measured $g^{(1,1)}$ never drops below a value of 50. Hence the noise gate enables the source to be operated at a higher coincidence count rate for a fixed amount of uncorrelated noise in the heralded output.


\section{Discussion}

Although noise gating can significantly improve certain aspects of the performance of a heralded single-photon source, it should be noted that it is not a complete solution to achieving the highest performance. Ultimately, performance will be limited by the fundamental thermal statistics of the generation process, which imposes a limit for a single source of 0.25 on the probability of delivering one pure heralded photon per mode. The only technique that can bypass that restriction is actively combining the output of several generation modes to create a multiplexed source~\cite{Migdall2002Tailoring-Single-Photon,Christ2012Limits-on-the-Deterministic}. Nevertheless, because noise gating increases the average pair-generation probability at which the sources can be operated for a given level of uncorrelated noise, it can reduce the number of sources that would be required in any realistic multiplexing scheme.

We envisage that improvements to our noise-gating scheme could be made by shrinking the switching window $\Delta t_{sw}$ around the arriving heralded photon to remove additional noise. More advanced noise gating could be carried out using a heralding detector with photon number resolution (PNR) capability to enable discrimination between single- and multi-pair emission. This could be achieved either with a superconducting PNR detector or a pseudo-PNR detector constructed from binary detectors~\cite{Achilles2003Fiber-assisted-detection-with, Fitch2003Photon_Number_Resoltuion}.

\section{Conclusion}

We have demonstrated that a low-loss fibre-integrated switch can improve the coincidence-to-accidental ratio of an all-fibre source of high-purity heralded single photons by an average factor of 7 by removing unheralded noise photons from the output. We are currently working on multiplexing this source in the temporal domain to  improve further the probability of delivering heralded single photons from the source.

\section*{Funding Information}
This work was funded by the UK EPSRC First Grant scheme (EP/K022407/1) and the EPSRC Quantum Technology Hub \textit{Networked Quantum Information Technologies} (EP/M013243/1).

\end{document}